# Suitability of FPS and DPS in NOMA for Real-Time and Non-Real Time Applications.

Moontasir Rafique
*Electrical and Electronic Engineering*
*Islamic University of Technology*
Gazipur, Bangladesh
moontasir@iut-dhaka.edu

Abdullah Alavi
*Electrical and Electronic Engineering*
*Islamic University of Technology*
Gazipur, Bangladesh
abdullahalavi@iut-dhaka.edu

Aadnan Farhad
*Electrical and Electronic Engineering*
*Islamic University of Technology*
Gazipur, Bangladesh
aadnanfarhad@iut-dhaka.edu

Mohammad T. Kawser
*Electrical and Electronic Engineering*
*Islamic University of Technology*
Gazipur, Bangladesh
kawser@iut-dhaka.edu

*Abstract*— **Non-Orthogonal Multiple Access (NOMA) is a popular solution for supporting a high number of users and along with significant bandwidth in 5G cellular communication. By using a technique called cooperative relaying, the same data is sent to all the users, and one user can relay data to the other. In order to provide enough power for the users, energy harvesting techniques have been introduced with Simultaneous Wireless Information and Power Transfer (SWIPT) coming to prominence in recent times. In this paper, analysis has been made comparing two different power allocation schemes in NOMA, Fixed Power allocation Scheme (FPS) and Dynamic Power allocation Scheme (DPS). The comparisons were made in terms of their performance and characteristics while undergoing SWIPT. It has been found that by using DPS, an almost 25% increase in peak spectral efficiency can be obtained compared to FPS. However, DPS suffers from a higher outage probability as the increase of power causes the signal bandwidth to drop below the target rate a significant number of times. Based on the detailed results, conclusions were drawn as to which power allocation coefficient scheme would be used in real-time and non-real time communication standards, respectively. The results suggest that for real-time communication, FPS is more suitable while for non-real time communication, DPS appears to work better than FPS.**

## I. INTRODUCTION

The dawn of the fourth industrial revolution has ushered in an era of massive avalanches of advancements and innovations in the field of technology, and with an ever-increasing number of mobile users, IoT devices, smart cities, and cloud computing, the demand for a faster, more versatile and reliable wireless network is at an all-time high. This calls for a proper operation and management of the wireless resources, and thus it is of utmost importance to ensure that clients have sufficient data rates for a particular task. In LTE, one of the goals of network operators was to minimize the inter-cell interference between the users while maintaining higher bandwidth. Thus, Orthogonal Frequency Division Multiple Access (OFDMA) was introduced which ensured that users were orthogonal to each other in terms of their cellular connectivity. In other words, no two users were allocated the same wireless resource. However, this resulted in a limited capacity of devices. Therefore, in the current day and age, where the needs of mobile users are increasing day by day, it became evident that a new scheme would be necessary where both the spectral efficiency and number of users supported were sufficiently higher. Thus, a new concept known as Non-Orthogonal Multiple Access (NOMA) was developed in 5G [1]. Here, the users did not maintain orthogonality, rather, the same resource was allocated for the users unlike OFDMA, and it was observed that both the spectral efficiency and number of supported devices were higher in the case of NOMA [2]. This could potentially play a pivotal role in upcoming generation communication standards. The idea behind NOMA is based on a technique which allows users to share the same resource with some certain conditions [3]. During transmission, it follows the procedure known as superposition coding [4], while in the case of reception, the receiver resorts to a process called Successive Interference Cancellation (SIC) [5] to decode its own signal from the original one. The numbering of the users are done in terms of their distances from the eNodeB. The farther the user, the more power is allocated to it. Naturally, more power is assigned to the far user than the near one. This is known as power allocation [6] and the power allocation procedure that has been discussed above is known as Fixed Power allocation Scheme (FPS) algorithm [7]. Here, the power allocation coefficients do not change with time. Alternatively, Dynamic Power allocation Scheme (DPS) [8] allows the power allocating coefficients to change with time.

In NOMA, there is provision for the near user to relay the data to the far user if there is an obstacle between the eNodeB and the far user. The process is referred to as cooperative relaying. This can only happen in NOMA as the near user carries data of the far user. However, the near user is a simple user equipment (UE) which has limited power supply as compared to the eNodeB. Therefore, considering the limited battery life of the UE, relaying information to the far user is an energy intensive process for the near user. To overcome this problem, a process known as wireless energy harvesting is used. In this technique, the UE divides the received signal power into two portions, one for energy harvesting, and the other for data decoding. There are mainly two ways of dividing the process, namely time switching and energy splitting. In this paper, energy splitting protocol is considered, more specifically, Simultaneous Wireless Information and Power Transfer (SWIPT) [9]. Here, the receiver uses a portion of the received signal to harvest energy and the remaining power is used to decode the incoming signal. Both the harvesting and decoding process is simultaneously completed in the same time frame. In the next time frame, the harvested energy is used to relay the signal to the far user.

In this paper, it has been shown how Fixed Power allocation Scheme (FPS) and Dynamic Power allocation Scheme (DPS) works differently in SWIPT. From that, a tradeoff has been established based on the two algorithms in real time and non-real time communication standard. The remainder of the paper has been organized as follows. In section II, the system model has been shown and elaborated. Next, the simulation model and its parameters have been discussed in section III. Finally, the results of the simulation have been discussed in section IV before presenting the conclusion in section V.

## II. SYSTEM MODEL

A network model is considered consisting of two user equipment (UE) and one eNodeB. The channel considered here is a Rayleigh fading channel. A barrier has been placed between the far user and the eNodeB which creates hindrance to direct information transfer, resulting in cooperative relaying procedure being undertaken. The distance from the far user to the eNodeB has been taken as exact double of the distance from near user to eNodeB for ease of experimenting. Fig. 1 illustrates the scenario which the simulation tries to emulate.

In NOMA, different power levels are allocated to the users corresponding to their distances from the eNodeB. This power allocation can be done in a number of manners. In this paper, Fixed Power Allocation (FPS) scheme and Dynamic Power Allocation (DPS) scheme.

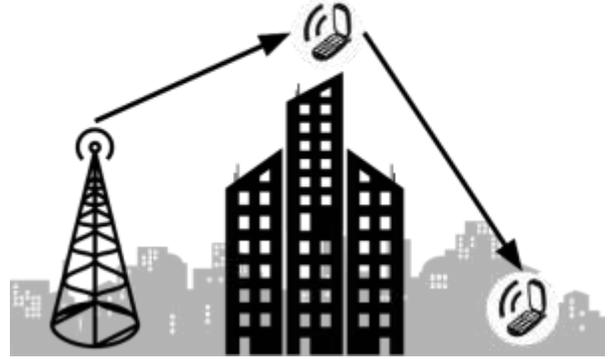

Fig. 1. Proposed Network Model comprising of two user equipment and one eNodeB.

### A. Fixed Power Allocation (FPS):

In this process, power allocation coefficients were fixed throughout the time. That is, they remained time invariant. The near user remained closer to the eNodeB compared to the far user and thus it has a better signal coming from the eNodeB. For this reason, the power allocation coefficient that is lower than the far user is allotted for the near user. The conditions for fixed power allocation are stated below:

$$a_n = 1 - a_f \quad (1)$$
$$a_f > a_n \quad (2)$$

Where, $\alpha_f$ signifies the power allocation coefficient for the far user; $\alpha_n$ signifies the power allocation coefficient for the near user [10]. The achievable rate in bps for both near user and far user are shown below,

$$R_{fu} = log_2(1 + \frac{|h_f|^2 P a_f}{|h_f|^2 P a_n + \sigma^2}) \quad (3)$$

$$R_{nu} = log_2(1 + \frac{|h_n|^2 P a_n}{\sigma^2}) \quad (4)$$

Where, $R_{nu}$ and $R_{fu}$ signifies the achievable rate in bps for the near and far user respectively. The term $\sigma^2$ signifies the total noise in the system. Again, $h_n$ and $h_f$ respectively denote the Rayleigh fading coefficients.

### B. Dynamic Power Allocation (DPS):

Dynamic Power allocation or Fair power allocation is a very effective form of power allocation where the coefficients are dynamic [11]. In case of FPS, the coefficients remain the same throughout the time. Because of this constraint, when the user is in a mobile state, there arises difficulty in assessing the SIC procedure, as mobility greatly puts an effect on interference as well as the channel quality. Thus, DPS allows flexibility in this regard where the values of coefficients are not limited to a certain value. In order to find the coefficients, we have to set a rate target for the UE to achieve, according to which the powers would be allocated. A target SINR value and rate value should be set in order to find the dynamic coefficients. The goal should be to keep the rate of the far user always higher than the target rate. That is,

$$R_{fu} \geq R^* \quad (5)$$

For allocating power, equation 1 and 2 has to be true. The target SINR (Signal-to-Interference-plus-Noise-ratio) rate, shown by *S*, where,

$$S = 2^{R^*} - 1 \quad (6)$$

For DPS, the near and far coefficients are shown below,

$$a_f = \frac{S(|h_f|^2 P + \sigma^2)}{|h_f|^2 P(1+S)} \quad (7)$$

$$a_n = 1 - a_f \quad (8)$$

Where, $h_f$ is the Rayleigh fading coefficient of the far channel, *S* is the Target SINR rate, and $\sigma^2$ denoting the noise power. *P* is the transmit power.

### C. Cooperative Relaying:

In NOMA, the eNodeB sends the same coded signal for all the receivers. That is, irrespective of the number of users, the eNodeB sends the same data differentiated by superposition coding to all the user equipment. The user nearer to the base station will have the information for the far user as well. The near user obtains its dedicated information by decoding the data of the far user. This provides an added advantage to NOMA, that is; in instances where the far user doesn't have a proper connection with the eNodeB, the near user has the capability to relay the information to the far user. This procedure is known as Cooperative Relaying [12]. By this process, more than one connection can be established between the eNodeB and the user equipment. For a two user NOMA scheme, a comparison is shown for the far user where cooperative relay procedure has been used in terms of outage probability and the total transmitted power.

In Fig. 2, comparison has been made between far users with cooperative relay and without cooperative relay phenomenon. It is evident from the graph that outage is lower when cooperative relay procedure is used.

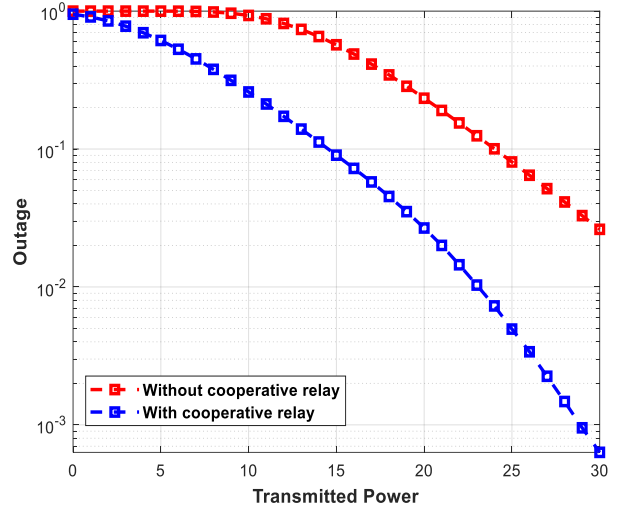

Fig. 2. Outage vs Transmitted power for the far user with and without cooperative relay.

### D. Radio Frequency Harvesting

There are different types of radio frequency harvesting procedures. In this paper, focus has been given to SWIPT (Simultaneous Wireless Information and Power Transfer) [13]. In a Rayleigh channel, there may lie barriers in between the base station and the far user. In a scenario like this where the far user has poor connection from the eNodeB, the near user may relay the information to the far user. This process is known as cooperative relaying. The method used in this paper is SWIPT; where a fraction of the received power is decoded, and the other part is harvested by the user [14]. For SWIPT, two time slots are allotted. The user which is near to the eNodeB, harvests some amount of power. And in the succeeding time slot, the rest of the power obtained is used for information decoding [15]. If the near user harvests ω amount of power in the first slot, for decoding information, the remaining fraction can be used, where Ω signifies the remaining power harvesting fraction [16].

$$\Omega = 1 - \omega \quad (9)$$

If $P_H$ is the total harvested power, it can be written according to the following equation:

$$P_H = P|h_n|^2 \zeta \omega \quad (10)$$

Where, ζ is the power harvesting efficiency, ω is the amount of power harvested in the first slot.

### III. SIMULATION AND RESULT ANALYSIS

The simulation was carried out using Log Normal Shadowing as the Path Loss Model. Since there is no direct line of sight signal, particularly between the far user and the eNodeB, Rayleigh Fading Channel has been used. For this particular simulation, the path loss exponent was set to 4. The bandwidth was set to 10 GHz with a target spectral efficiency of 1 bps/Hz. As we are considering a simple case of a near user with a far user, the UE number was set to 2. The power allocation coefficient for the near user was set to 0.2 while the far user's was set to 0.8. In addition, the power harvesting factor for the near user was set to 0.7. The transmission power from the base station was repeated from 0 to 30 dBm with 5 dBm increments. The simulation was modeled to keep the far user behind an obstacle. The distance from the base station to the near user and the far user was 10 and 20 meters, respectively. Table 1 summarizes the simulation parameters that were used for this simulation.

Table 1 Simulation Parameters.

| Parameters of Simulation | |
|---|---|
| Bandwidth | 10 GHz |
| Target Spectral Efficiency | 1 bps/Hz |
| Number of UE | 2 |
| BS power | (5,10,15,20,25,30) dbm |
| Path loss exponent | 4 |
| Fixed power allocation coefficient of near user | 0.2 |
| Fixed power allocation coefficient of far user | 0.8 |
| Near user Power Harvesting Factor | 0.7 |
| Near user from eNodeB distance | 10m |
| Far user from eNodeB distance | 20m |

In this section, different results based on the simulation model have been analyzed. In Fig. 3, we can see the average spectral efficiency of the near and far user in terms of the transmitted power level for both DPS (Dynamic power Allocation Scheme) and FPS (Fixed Power Allocation Scheme). From Fig. 3, it is evident that the far users maintain a similar trend in both DPS and FPS. That is, the Average Spectral Efficiency remains unchanged for the far user in both dynamic power allocation and fixed power allocation schemes. But, in the case of a nearby user, it can be seen that they vary hugely. In low power level (from the graph, 0-10 dBm), DPS gives better spectral efficiency than FPS. This trend remains the same up to around 12 dBm for DPS, after which it starts to deteriorate. After a specific power level (around 20 dBm from the graph), FPS provides better spectral efficiency than DPS and keeps on increasing. So, for a low power level, it is evident that DPS provides better spectral efficiency than FPS. In terms of percentage, it is seen that for 0-12 dBm region transmit power, DPS provides about 150% of higher bandwidth than FPS. In the middle region (22-30 dBm in the graph), FPS leads DPS in terms of bandwidth. Finally, DPS becomes higher than FPS again in high power regions (greater than 30 dBm).

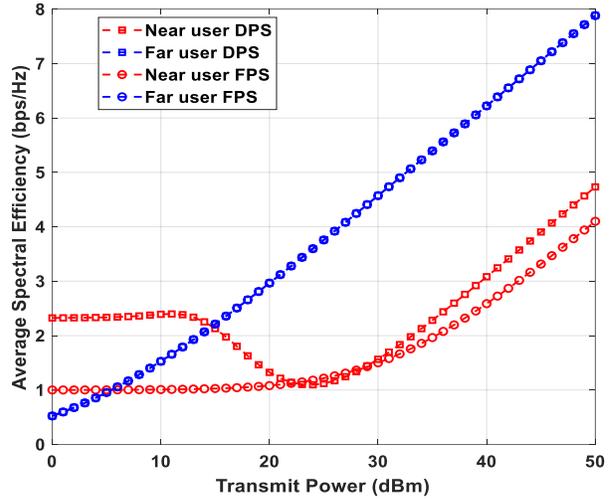

Fig. 3. Average Spectral Efficiency vs Transmitted Power for both DPS and FPS.

In Fig. 4, a comparative analysis between power outage and transmitted power has been made. It is seen that, for outage too, the far user maintains a similar fashion for both FPS and DPS. That is, its trend remains similar, it decreases as the transmitted power is increased.

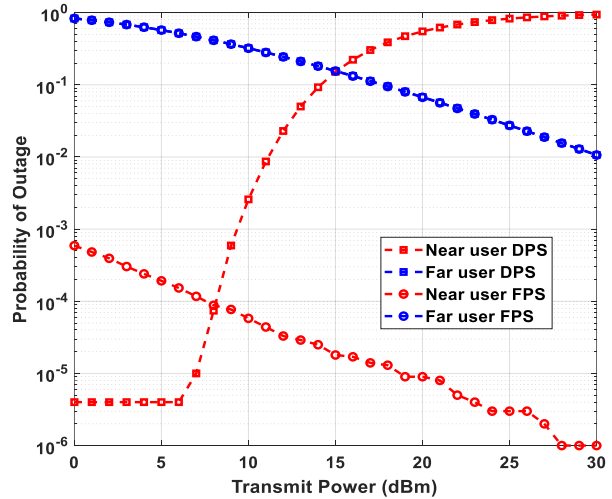

Fig. 4. Outage vs Transmitted Power for both DPS and FPS.

So, it can be understood that the power allocation scheme does not have any significant effect on the far user both in terms of spectral efficiency and outage. In the case of nearby users, we can see they vary hugely. For DPS, the near user outage keeps on increasing as the power is increased to its maximum value; that is the outage increases as power is increased continuously after maintaining a constant value (up to 6 dBm from the graph). And, in case of FPS, the near user has higher outage than the far user in low power level (0-6 dBm from the graph). But as the power is increased, the outage decreases and reaches a minimum value when the power is the maximum. It can be inferred that for lower power level, near users have higher outage in FPS. As we increase the power level, the outage of FPS decreases in comparison to DPS.

In Fig. 6 and Fig. 5, comparison has been made between the average spectral efficiency and Channel Realization Instances between the near and far user for both DPS and FPS. A reference rate has been set (1bps/Hz).

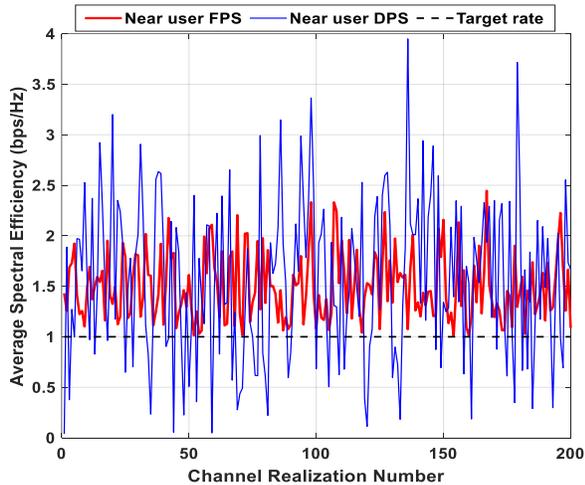

Fig. 6. Average Spectral Efficiency vs Channel Realization Number for Near User.

For near user FPS, in Fig. 6, the spectral efficiency always remains above the reference rate that had been set earlier (in the graph, colored in red). And in case of DPS, it can be seen that the rate falls drastically below the average rate for some channel realization instances. But it is also evident that the highest spectral efficiency is obtained for DPS scenario (in the graph, colored in blue) which is almost greater than 3 bps/Hz compared to 2.5 bps/Hz for FPS (almost 25% increase in Peak Value). That is, on average, FPS has higher average spectral efficiency peaks than DPS, even though it falls below the reference rate several times. This signifies that there exists severe outage in DPS when compared to FPS scenario.

In Fig. 5, in the case of far user, it is apparent that the average spectral efficiency of both DPS and FPS bear a similar trend in terms of the channel realization number. Moreover, they exhibit the same property throughout the graph while decreasing below the target rate only once for both DPS and FPS. Thus, it shows that for the far user, DPS and FPS bear similar characteristics irrespective of the power allocation scheme chosen.

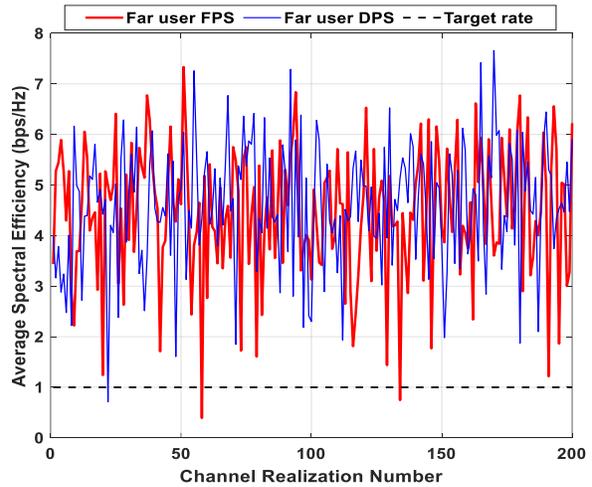

Fig. 5. Average Spectral Efficiency vs Channel Realization Number for Far User.

## IV. CONCLUSION

In this paper, fixed and dynamic power allocation algorithms were compared in SWIPT method. It was seen that even though as per our network model, the far user exhibits similar characteristics for both FPS and DPS method, the near user shows different properties for each. In FPS, the near user never goes below the targeted bandwidth, as a result, proper signal quality is always ensured. But, in the case of DPS, it has been observed that the signal falls below the target rate a number of times, inferring that the target bandwidth is not always maintained in DPS. Moreover, it has also been seen that the peak average spectral efficiency in FPS is almost 25% more than that of DPS; establishing the fact that it provides higher bandwidth compared to DPS in average. Furthermore, it is seen that for low power, DPS provides substantially higher bandwidth compared to FPS. In high power range, DPS is again better in terms of spectral efficiency compared to FPS for the near user. Thus, based on the results, it can be suggested that for real time communication, FPS appears to be more suitable than DPS. In contrast, for non-real time standards, DPS appears to work better than FPS.

## V. Application And Discussion

Communication standards can be divided into two types based on time constraints, namely real time, and non-real time. In real time processes, it is of absolute necessity that the target rate is met, i.e., the data rate should always be greater than the threshold value to maintain good quality. In contrast, for non-real time standards, it is not mandatory for the data rate to meet the target demand, rather, how high the data rate is what matters significantly. These kinds of processes do not occur instantly like real time as there are no time constraints involved in this procedure. From the results obtained, it is seen that the far user exhibits similar property both in FPS and DPS scenarios. However, speaking of the near user, they results vary quite considerably. In DPS, the data rate will go below the target rate a number of times, thus establishing the fact that it cannot be used in the case of real time communication. Whereas, in FPS, it can be seen that it always remains above the target rate, suggesting that real time communication can be supported using this scheme. Additionally, the peak average spectral efficiency of DPS is almost 25% greater on average than FPS. Thus, it can be said that for non-real time communication standards where higher data rate is preferred over time constraints, DPS appears to be more suitable compared to FPS.